\documentclass[cameraready]{Interspeech}

\title{Instantaneous Pitch Estimation via\\Wave-U-Net-Based Fundamental Waveform Enhancement}

\author[affiliation={1}, orcid=0009-0007-0682-1546, equalcontribution]{Junya}{Koguchi}
\author[affiliation={1}, orcid=0000-0002-8347-5604, equalcontribution, correspondingauthor]{Tomoki}{Koriyama}

\address{
    $^1$ CyberAgent, Japan
}

\email{koguchi\_junya@cyberagent.co.jp, koriyama\_tomoki@cyberagent.co.jp}


\usepackage{comment}
\usepackage{wasysym}
\usepackage{booktabs}
\usepackage{multirow}
\usepackage{cite}

\newcommand{\Fig}[1]{\textbf{Figure.~\ref{fig:#1}}} 
\newcommand{\Table}[1]{\textbf{Table~\ref{tb:#1}}} 
\newcommand{\Eq}[1]{Eq.~(\ref{eq:#1})} 
\newcommand{\Subsec}[1]{\textbf{Section~\ref{subsec:#1}}} 

\newcommand{\revise}[1]{\textcolor{black}{#1}}

\begin{document}

\maketitle

\begin{abstract}
    Instantaneous pitch estimation plays an important role in analyzing steep pitch variations such as speech prosody and singing techniques. Conventional approaches estimate instantaneous frequency after isolating the fundamental waveform from signals that contain harmonics and noise, which makes the accuracy sensitive to imperfect fundamental filtering. In this study, we formulate fundamental waveform filtering as a speech enhancement problem. Specifically, we train a Wave-U-Net model to extract a fundamental waveform from an input speech signal. The instantaneous pitch is then obtained by computing the instantaneous frequency from the analytic signal of the estimated fundamental waveform. Experimental results show that the proposed method outperforms conventional deterministic approaches and provides accurate and robust instantaneous pitch estimation across diverse domains, including speech, singing voice, musical instruments, and degraded speech signals.
\end{abstract}

\section{Introduction}
The fundamental frequency of a speech signal ($f_\mathrm{o}$~\footnote{We use $f_\mathrm{o}$ as recommended by Acoustical Society of America. The notation F0 is misleading since it may suggest the “0th formant,” and it can be confused either with the lowest resonance frequency of audio devices or with the lowest musical note F.}) roughly corresponds to perceived pitch. Therefore, accurate $f_\mathrm{o}$ estimation algorithms are a core technology for analyzing prosody and singing techniques.
Many $f_\mathrm{o}$ estimators split a signal into short-time frames and estimate pitch for each frame as a discrete value of period or frequency.
Because such frame-wise estimation assumes that $f_\mathrm{o}$ is stationary within each frame, it tends to produce unnatural discontinuities when $f_\mathrm{o}$ changes continuously, as in vibrato or chirp signals~\cite{azarov2012irapt}. It also becomes difficult to track abrupt changes~\cite{Kawahara2012highspeed}.
\begin{figure}[t]
  \centering
  \includegraphics[width=\columnwidth]{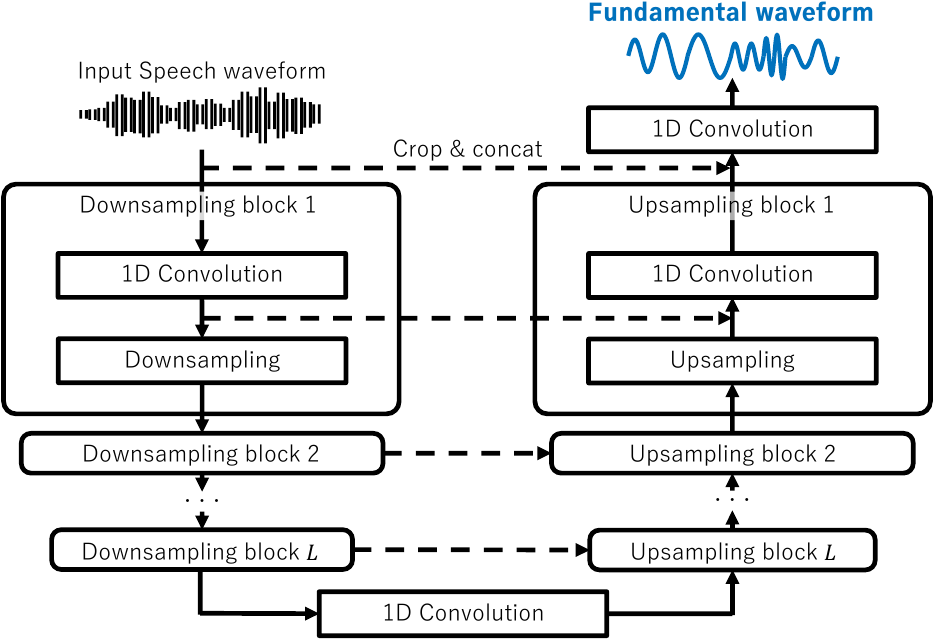}
  \caption{Overview of the proposed Wave-U-Net-based fundamental waveform filtering.}
  \label{fig:overview}
  \vspace{-20pt}
\end{figure}

To address this issue, instantaneous pitch estimation (\revise{IPE}) estimates a continuous $f_\mathrm{o}$ by directly computing the instantaneous frequency of the fundamental component~\cite{Abe1995instpitch, azarov2012irapt, azarov2016halcyon, Kawahara2017ninjal1, Kawahara2017ninjal2, blok2021ife}.
Instantaneous frequency is defined as the time derivative of the phase of an analytic signal, which corresponds to estimating the modulation frequency of a single sinusoidal component.
If we can extract only the sinusoidal component corresponding to $f_\mathrm{o}$ from a speech signal (hereafter referred to as the fundamental waveform), we can obtain a continuous $f_\mathrm{o}$ trajectory by computing its instantaneous frequency.
However, real speech is a multi-component signal that contains harmonic structures and noise components, and instantaneous frequency is not uniquely defined in general.
Therefore, a key challenge in \revise{IPE} is to select, with high accuracy, the channel that contains the fundamental waveform from the outputs of a complex filterbank.
In conventional signal-processing-based approaches, each channel after band decomposition is scored for ``fundamental-likeness'' using criteria such as autocorrelation~\cite{azarov2012irapt} and standard deviation~\cite{Kawahara2017ninjal1, Kawahara2012highspeed}.
While these deterministic approaches have clear theoretical motivations, they are often brittle under signals from domains that were not assumed at design time or under unseen noise conditions.

Wave-U-Net, which directly takes and outputs time-domain waveforms, was proposed for source separation~\cite{Stoller2018waveunet}. It has also been applied to speech enhancement~\cite{Macartney2018wunse}, and it is known to extract a desired waveform component from a mixture with high accuracy.
In addition, a Wave-U-Net-based method that directly estimates pitch marks from waveforms has been reported~\cite{nam2023waveupm}, which suggests that Wave-U-Net is also effective for learning periodic structures in the time domain.

In this study, we formulate fundamental waveform filtering for \revise{IPE} as a specialized speech enhancement problem, and we utilize Wave-U-Net's strong waveform enhancement capability.
Specifically, we train Wave-U-Net to extract only the fundamental waveform from an input speech signal (\Fig{overview}).
This design allows us to obtain instantaneous pitch directly from the enhanced fundamental waveform without channel selection. It also aims to robustly enhance the fundamental waveform even when the input speech contains noise and strong harmonic components.

In our experimental evaluation, we confirm that the proposed method achieves higher estimation accuracy than \revise{conventional deterministic approaches based on signal processing}.
We also show that the proposed method yields continuous and smooth estimates even for signals with abrupt pitch changes or strong frequency modulation, which demonstrates the effectiveness and robustness of the proposed approach.

\section{Related works}
\subsection{Instantaneous pitch estimation}
The instantaneous frequency of an analytic signal is defined as the time derivative of its phase.
In practice, since a speech signal $x_\mathrm{sp}$ with time index $t$ is \revise{a real value} and has a harmonic structure, we first extract the analytic signal of the fundamental waveform $z_\mathrm{fund}$ using a complex bandpass filter and then compute the instantaneous frequency:
\begin{align}
        z_\mathrm{fund}(t) &= h(t; f_\mathrm{bw}, f_\mathrm{c})*x_\mathrm{sp}(t),\\
        h(t;  f_\mathrm{bw}, f_\mathrm{c}) &= 2 \frac{\sin\left(2 \pi f_\mathrm{bw} t\right)}{\pi t} \, w(t) \, \exp(j2\pi f_\mathrm{c} t). \label{eq:filter}
\end{align}
Here, $*$ denotes convolution, $w(t)$ denotes a window function, $f_\mathrm{bw}$ denotes the filter bandwidth, and $f_\mathrm{c}$ denotes the center frequency.

\subsection{Channel selection}
IRAPT~\cite{azarov2012irapt} constructs an instantaneous autocorrelation-like function from instantaneous amplitude and instantaneous frequency estimated in each band. It then tracks period candidates using dynamic programming to determine the fundamental period.
Halcyon~\cite{azarov2016halcyon} adaptively changes the filter bandwidth and analysis length for each candidate frequency using multi-rate sampling, and it evaluates period candidates based on the instantaneous harmonic parameters estimated at each step.
Both frameworks indirectly select the fundamental frequency based on periodicity measures that integrate harmonic information across multiple bands.
NINJAL~\cite{Kawahara2017ninjal1, Kawahara2017ninjal2} estimates the reliability of periodicity by combining band-wise residuals and deviations of phase derivatives, and it extracts fundamental frequency candidates from the minima of the resulting measure.

\section{Proposed method}
\subsection{Wave-U-Net-based fundamental waveform filtering}
Wave-U-Net~\cite{Stoller2018waveunet} is a U-Net-type~\cite{ronneberger2015unet} DNN consisting of an encoder and a decoder.
Wave-U-Net $\mathcal{F}_\mathrm{WUN}$ takes a speech waveform $x_\mathrm{sp}(t)$ as input and is trained as a regression model that directly estimates the fundamental \revise{waveform} $x_\mathrm{fund}(t)$:
\begin{equation}
    \hat{x}_\mathrm{fund}(t) = \mathcal{F}_\mathrm{WUN}(x_\mathrm{sp}(t)).
\end{equation}
We define the residual component as $x_\mathrm{res}(t)=x_\mathrm{sp}(t)-x_\mathrm{fund}(t)$ and $\hat{x}_\mathrm{res}(t)=x_\mathrm{sp}(t)-\hat{x}_\mathrm{fund}(t)$, which contains harmonic components and noise.
We train the model using the mean absolute error (MAE) so that the fundamental and residual components are correctly separated:
\begin{equation}
    \mathcal{L}_\mathrm{wave}=\mathbb{E} \left[ \lvert x_\mathrm{fund}(t)-\hat{x}_\mathrm{fund}(t)\rvert +  \lvert x_\mathrm{res}(t)-\hat{x}_\mathrm{res}(t)\rvert \right].
\end{equation}
Here, $\mathbb{E}[\cdot]$ denotes expectation.
By imposing a loss term on the residual as well, we encourage the model to satisfy mixture consistency.
We use MAE rather than mean squared error because we expect MAE to reduce excessive smoothing and to preserve fine waveform structures.

To stabilize training, we additionally introduce a loss term on the instantaneous frequency $f(t)$:
\begin{equation}
    \mathcal{L}_{\mathrm{IF}}= \mathbb{E} \left[m(t)\lvert\hat{f}_\mathrm{fund}(t)-f_\mathrm{fund}(t)\rvert \right].
\end{equation}
$f_\mathrm{fund}(t)$ and $\hat{f}_\mathrm{fund}(t)$ are calculated from $x_\mathrm{fund}$ and $\hat{x}_\mathrm{fund}$, respectively.
The mask $m(t)$ suppresses gradients in regions where instantaneous frequency is unstable, and it is defined as
\begin{align}
    m(t) =
     \begin{cases}
     1, & \mathrm{if}~ |z_\mathrm{fund}(t)| > \epsilon_\mathrm{amp} , \\
     0, & \mathrm{otherwise}.
     \end{cases}
\end{align}
Since instantaneous frequency is unstable in silent or unvoiced consonant regions, we evaluate errors only in segments where the instantaneous amplitude exceeds a preset threshold $\epsilon_\mathrm{amp}$.
We compute the instantaneous frequency using \Eq{filter} with $f_\mathrm{bw} = (f_\mathrm{ceil} - f_\mathrm{floor})/2$ and $f_\mathrm{c} = (f_\mathrm{ceil} + f_\mathrm{floor})/2$, where $f_\mathrm{floor}$ and $f_\mathrm{ceil}$ are the lower and upper bounds of candidate $f_\mathrm{o}$ values.

The final loss function is
\begin{equation}
    \mathcal{L}=\mathcal{L}_{\mathrm{wave}}
                +\lambda \mathcal{L}_{\mathrm{IF}}.
\end{equation}
Here, $\lambda$ is a weight that balances $\mathcal{L}_{\mathrm{wave}}$ and $\mathcal{L}_{\mathrm{IF}}$.

\revise{The overall architecture follows the standard Wave-U-Net framework~\cite{Macartney2018wunse}. The only difference is that, instead of enhanced speech waveform, our model is trained to estimate the fundamental waveform.}
Each encoder and decoder consist of $L$ downsampling (DS) blocks and $L$ upsampling (US) blocks, followed by a bottleneck layer and an output 1D convolutional layer (1D CNN).
Each DS block stacks a 1D CNN layer with a nonlinear function and a downsampling layer that discards every other feature sample and halves the temporal resolution.
Each US block stacks an upsampling layer and a 1D CNN layer with a nonlinearity.
In addition, the output feature map of each DS block is passed to the corresponding US block via skip connections.
Since convolution slightly changes the temporal length of feature maps, we crop the center part of the encoder feature map to match the decoder feature-map length, and then we concatenate them along the channel axis.
This design allows the decoder to use both abstract features from the bottleneck and high-resolution local information retained in shallow encoder layers.
As a result, the network can integrate context information over a wide receptive field while preserving temporally precise waveform structures.
For upsampling, we first restore intermediate values by interpolation and then apply convolution to double the temporal resolution.
We use interpolation-based upsampling instead of transposed convolution that inserts zeros between samples, which prevents aliasing artifacts caused by zeros.
Except for the output layer that uses a tanh nonlinearity to constrain the output range to $[-1, 1]$, we use Leaky ReLU~\cite{maas2013leakyrelu} as the nonlinear function throughout the network.
Repeated downsampling increases the receptive field and enables the model to capture long-range temporal dependencies.
At the same time, skip connections preserve high-resolution information from shallow layers, which improves the reconstruction accuracy of fine waveform shapes.

\section{Experimental evaluation}
\subsection{Dataset and preprocess}
\label{subsec:dataset}
We used the following datasets for training and evaluation:
speech (Bagshaw~\cite{bagshaw1993pitch}, Keele~\cite{plante1995keele}, CMU ARCTIC~\cite{kominek2004cmu-arctic}, PTDB-TUG~\cite{pirker2011ptdb-tug}, MOCHA-TIMIT~\cite{wrench1999mocha-timit}),
singing voice (MIR-1K~\cite{hsu2010mir-1k}),
and musical instruments (MDB-stem-synth~\cite{salamon2017mdb-stem-synth}).
\revise{All signals were resampled to 16~kHz and 16~bit. For a subset of the data, the provided ground truth contained many errors.
We therefore denoised the accompanying electroglottograph (EGG) waveform~\cite{rx11} and extracted $f_\mathrm{o}$ using the voting method~\cite{koguchi2026voting}.}
\revise{This dataset consists of 42 speakers, 19 singers and 25 instruments totaling 20.67 hours of audio. Training, evaluation and test sets are strictly disjoint at the utterance level across all datasets. No speaker/singer/instrument overlap occurs between splits.}

We computed the ground-truth fundamental waveform by complex filtering centered at ground-truth $f_\mathrm{o}$.
Since $f_\mathrm{c}$ is known in this setting, we designed the filter based on \Eq{filter} as
\begin{align}
    h(t; \revise{f}_\mathrm{c}) &= w_\mathrm{Gauss}(t; G_\mathrm{nbhd}) \exp(\revise{j 2\pi  f_\mathrm{c} t)},\\
    W_\mathrm{Gauss}(f; G_\mathrm{nbhd}) &= \exp\left(\frac{\ln(10)\,G_\mathrm{nbhd}}{20}\left(\frac{f}{f_\mathrm{c}}\right)^2
\right).
\end{align}
Here, $w_\mathrm{Gauss}$ and $W_\mathrm{Gauss}$ denote Gaussian windows in the time and frequency domains, respectively.
$G_\mathrm{nbhd}$ is the gain at the locations of harmonics adjacent to $f_\mathrm{o}$, and we set it to $-100$~dB~\cite{Kawahara2022tool}.

To improve robustness to noise, we added noise signals from NOISEX92~\cite{varga1993noisex92} and QUT-NOISE~\cite{dean2010qutnoise} to the speech signals.
Noise was added with probability 30\%, and the noise samples were chosen at random.
The signal-to-noise ratio (SNR) was randomly chosen between 0 and 30~dB.

\subsection{Training setups}
The number of layers for both DS and US blocks was set to $L=6$.
The loss weight was set to $\lambda=5.0$, which we chose a value that yields stable training.
For optimization, we used the RAdam optimizer~\cite{Liu2020radam} with ScheduleFree~\cite{defazio2024schedulefree}
($\beta_1=0.9$, $\beta_2 = 0.999$, learning rate is $1.0 \times 10^{-4}$).
We trained for 30 epochs with a batch size of 16.

For each batch, we randomly cropped a segment of 4096 samples from the full waveform, and Wave-U-Net takes and outputs a waveform of the same length.
For instantaneous frequency computation, we set $f_\mathrm{floor}=40$ and $f_\mathrm{ceil}=2000$.
$\epsilon_\mathrm{amp}$ is set to the value that corresponds to a gain of $-100$~dB.

\subsection{Evaluation metrics}
\subsubsection{Frequency error}
We used cent error and raw pitch accuracy (RPA) as an error metric for instantaneous pitch.
We restricted evaluation to segments where the reference $f(t)$ exists and the analytic-signal amplitude defined in \Subsec{dataset} is sufficiently large.

To evaluate noise robustness, we computed RPA on the test speech and noise pairs described in \Subsec{dataset} under clean conditions and under additive noise at SNR conditions of $(\infty, 30, 20, 10, 0)$~dB.

\subsubsection{Response to frequency modulation}
To evaluate tracking performance for time-varying pitch, we investigate the response to frequency modulation.
Specifically, we used a response measurement method~\cite{Kawahara2022tool} based on CAPRICEP~\cite{kawahara2021capricep}.
For an input frequency that adds a modulation component $f_\mathrm{m}(t)$ to a carrier frequency $f_\mathrm{c}$, CAPRICEP decomposes \revise{the estimator response $\hat{f}(t)$} into a linear time-invariant (LTI) component $H_\mathrm{LTI}$, a nonlinear time-invariant component $f_\mathrm{nonLTI}$, and a \revise{random and time-varying component $f_\mathrm{TV-rand}$} as
\begin{align}
    f(t) &= f_\mathrm{c}+f_\mathrm{m}(t),\\
    \hat{f}(t) &= f_\mathrm{c}+H_\mathrm{LTI}*f_\mathrm{m}(t) + f_\mathrm{nonLTI}(t) + f_\mathrm{TV-rand}(t).
\end{align}
We visualized tracking performance by varying the modulation frequency for a test signal that simulates the Japanese vowel $/a/$ with $f_\mathrm{c}=240$~Hz and a modulation extent of 50~cents.
As in the RPA evaluation, we also evaluated the same test signal with additive white noise at 30~dB SNR.

\subsection{Methods}
For comparison, we used existing methods based on \revise{IPE},
\textbf{IRAPT}~\cite{azarov2012irapt},
\textbf{Halcyon}~\cite{azarov2016halcyon},
and \textbf{NINJAL}~\cite{Kawahara2017ninjal1, Kawahara2017ninjal2}.
We selected these methods based on accuracy and the availability of public implementations.

\subsection{Result and discussion}
\begin{figure*}[t]
  \centering
  \includegraphics[width=\linewidth]{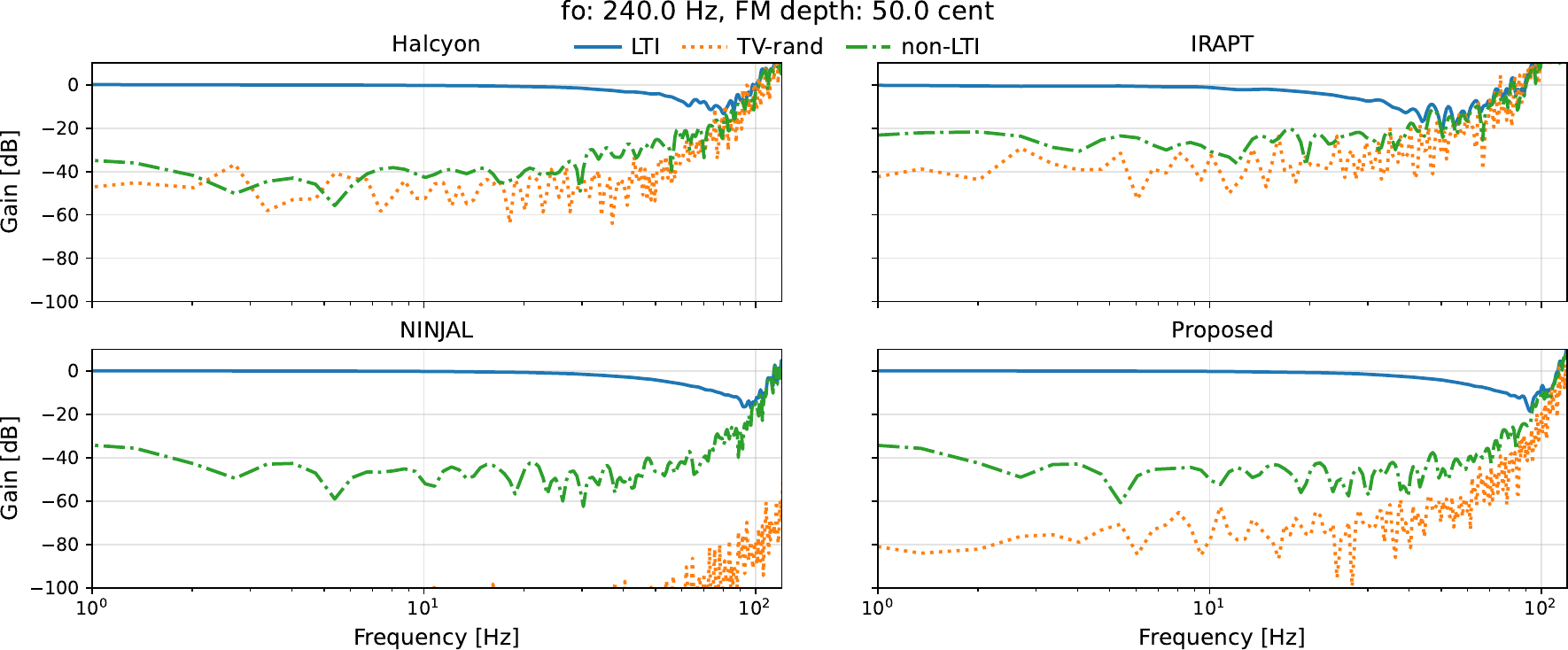}
  \vspace{-15pt}
  \caption{Frequency responses of each estimator under clean conditions.}
  \label{fig:capricep_clean}
\end{figure*}
\begin{figure*}[t]
  \centering
  \includegraphics[width=\linewidth]{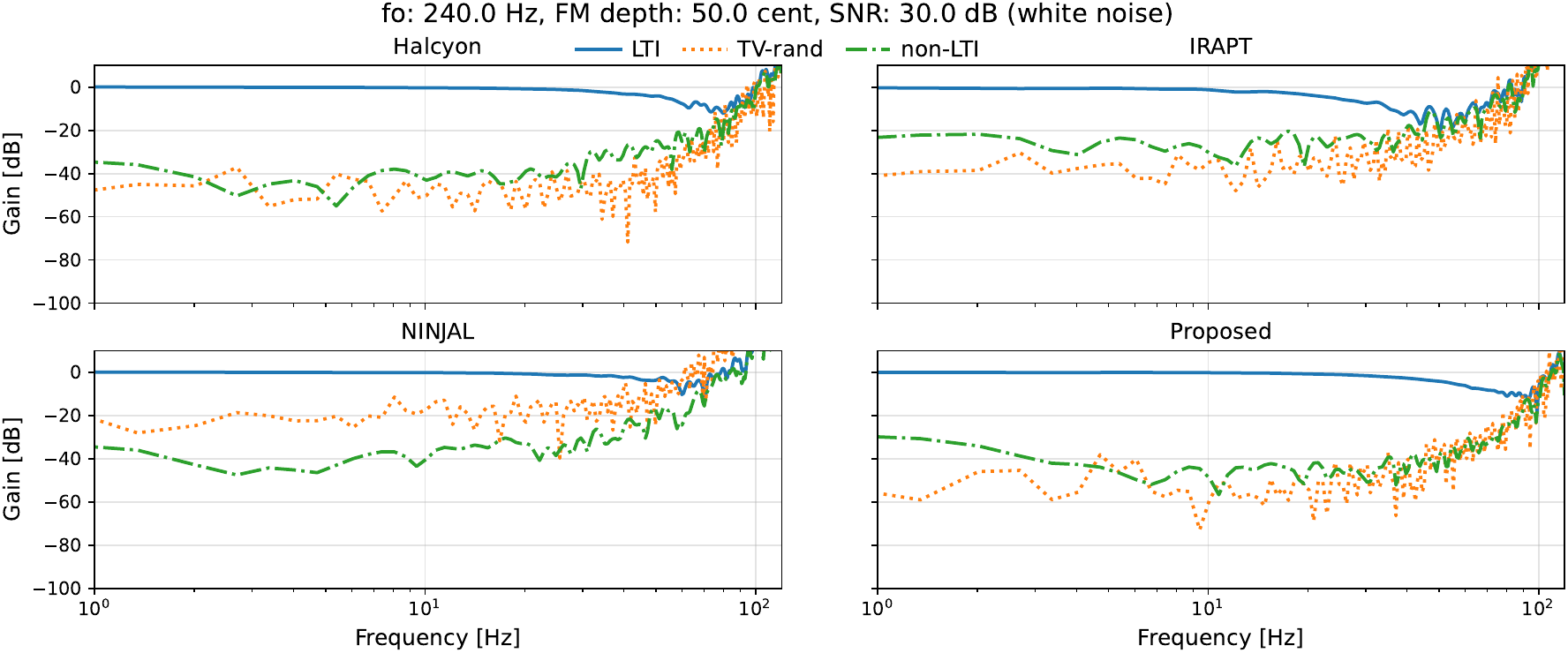}
  \vspace{-15pt}
  \caption{Frequency responses of each estimator under noisy conditions.}
  \label{fig:capricep_noise}
  \vspace{-15pt}
\end{figure*}

\begin{table}[tb]
\centering
\caption{Comparison of RPA for each \revise{IPE} method under clean conditions.}
\vspace{-10pt}
\label{tb:rpa_clean}
\begin{tabular}{lcccc}
\hline
\multirow{2}{*}{Method}
 & \multirow{2}{*}{$\Delta \cent$ $\downarrow$}
 & \multicolumn{3}{c}{$\mathrm{RPA}$ $\uparrow$} \\
\cmidrule(lr){3-5}
 &  & 5 & 25 & 50 \\
\hline
IRAPT    & $71.69 \pm 267.62$ & 17.35 & 65.54 & 83.84 \\
Halcyon  & $2.45 \pm 213.96$ & 37.01 & 78.57 & 86.80 \\
NINJAL & $30.41 \pm 564.64$ & 38.22 & 76.73 & 84.87 \\
Proposed & $28.23 \pm 133.98$ & 38.50 & 79.11 & 88.47 \\
\hline
\end{tabular}
\end{table}

\begin{table}[tb]
\centering
\caption{RPA under additive noise (threshold 50 cents).}
\vspace{-10pt}
\label{tb:rpa_noise}
\resizebox{\linewidth}{!}{%
\begin{tabular}{lccccc}
\hline
\multirow{2}{*}{Method $\backslash$ SNR [dB]} & \multicolumn{5}{c}{$\mathrm{RPA}_{50}$$ \uparrow$} \\
\cmidrule(lr){2-6}
& $\infty$ & 30 & 20 & 10 & 0 \\
\hline
IRAPT    & 83.84 & 83.98 & 84.05 & 84.30 & 81.41 \\
Halcyon  & 86.80 & 87.62 & 87.45 & 85.66 & 76.30 \\
NINJAL & 84.87 & 84.81 & 84.10 & 82.02 & 62.35 \\
Proposed & 88.47 & 88.28 & 88.61 & 87.46 & 86.40 \\
\hline
\end{tabular}%
}
\vspace{-10pt}
\end{table}

\noindent
\Table{rpa_clean} and \Table{rpa_noise} show the RPA results under clean and noisy conditions, respectively.
Under both conditions, the proposed method achieved the highest RPA.
This result suggests that Wave-U-Net stably extracts the fundamental waveform from both clean and noisy speech.
Under clean conditions, Halcyon achieved the smallest mean $\Delta \cent$, but it had a large standard deviation, and its RPA did not exceed the proposed method.
This trend likely reflects frequent outliers in segments with locally abrupt pitch changes, such as onsets and regions near voiced consonants.
The proposed method had a slightly larger mean error, but it had a smaller standard deviation and higher RPA even under the strict 5~cent threshold.
This result suggests that the estimated trajectory is more continuous and tracks abrupt changes better.

\Fig{capricep_clean} and \Fig{capricep_noise} show the modulation-response plots.
\revise{The horizontal axis denotes the modulation frequency of the input pitch trajectory. Lower TV-rand and non-LTI responses indicate more stable IPE.}
Under clean conditions, NINJAL showed the smallest random response and maintained high tracking performance even as the modulation frequency increased.
The random response of the proposed method was smaller than those of IRAPT and Halcyon, but it was larger than that of NINJAL.
This gap suggests that a small amount of harmonic components may remain in the estimated fundamental waveform.
One possible cause is that aliasing components introduced by Wave-U-Net downsampling affect phase differences. This effect can appear as random and nonlinear responses.
This issue may be mitigated by adding anti-aliasing downsampling with perfect reconstruction guarantees, for example by introducing discrete wavelet transforms~\cite{nakamura2021mrdla}.

Under noisy conditions (white noise at 20~dB), we observed a notable increase in the random component $f_\mathrm{rand}$ for NINJAL.
Instantaneous frequency assumes that the signal is locally a single sinusoidal component, and the phase difference becomes sensitive to random fluctuations when noise dominates.
This sensitivity can produce spiky errors and tracking failures in instantaneous pitch estimates.
Consistent with this observation, \Table{rpa_noise} also shows a large drop in RPA at SNR 0~dB.
In contrast, the proposed method showed relatively small changes in $H_\mathrm{LTI}$ and $f_\mathrm{rand}$ under noisy conditions.
This behavior may be attributed to two factors: noise augmentation during training encouraged Wave-U-Net to suppress noise components unrelated to the fundamental waveform, and $\mathcal{L}_\mathrm{IF}$ encouraged stable phase evolution of the extracted fundamental waveform.
It has been reported that DNN-based $f_\mathrm{o}$ estimation methods are generally robust to noise~\cite{terashima2025slash}, and our results are consistent with this trend.

In practice, NINJAL and Halcyon can be advantageous when one aims to analyze strong modulation precisely in voiced segments dominated by periodic components, such as vowel-only utterances and vibrato.
On the other hand, for general speech that includes consonants or for recordings in noisy environments, the proposed method is likely to be effective.
The degree to which aperiodic components, such as noise and breath noise, dominate relative to harmonic components can serve as a criterion for method selection.

\section{Conclusion}
We proposed Wave-U-Net-based fundamental waveform filtering for robust \revise{IPE} from speech.
By directly extracting the fundamental waveform using a DNN before \revise{IPE}, the proposed method omits the need for channel selection in conventional complex filterbanks and targets robustness to mixtures of noise and harmonics.
Experimental results showed that the proposed method achieved higher accuracy than existing instantaneous pitch estimatiors, and it was particularly robust under strong noise conditions.

Future work includes architectural improvements to reduce nonlinear components, such as revisiting downsampling designs with anti-aliasing, and applying the proposed method to speech and singing voice analysis in practical settings.

\section{8. Use of generative AI tools}
\revise{ChatGPT was used to aid editing and polishing this manuscript.}

\bibliographystyle{IEEEtran}
\bibliography{bib/interspeech2026}

\end{document}